\documentclass[amssymb,prl,twocolumn,showpacs]{revtex4}
\usepackage{amsmath}
\usepackage{graphicx}
\usepackage{verbatim}
\usepackage{graphics}
\usepackage{units}
\usepackage{ulem}
\usepackage{color}
\newcommand{\bea}{\begin{eqnarray}}
\newcommand{\eea}{\end{eqnarray}}
\newcommand{\be}{\begin{equation}}
\newcommand{\ee}{\end{equation}}

\begin{document}

%{\!\ Phys. Rev. Lett. {\bf XX}, xxxxxx (2014)}

\title{
\vspace*{-1.25cm}
\textnormal{{\small \flushright PHYSICAL REVIEW LETTERS {\bf 112}, 176803 (2014)}}\\
\vspace*{-0.2cm}
\rule[0.1cm]{18cm}{0.02cm}\\
%\ \\
\vspace*{0.285cm}
Long-range spin transfer in triple quantum dots
}
\author{R. S\'anchez$^1$, G. Granger$^2$, L. Gaudreau$^{2,3}$, A. Kam$^2$, M. Pioro-Ladri\`ere$^3$, S.~A.~Studenikin$^2$, P. Zawadzki$^2$, A. S. Sachrajda$^2$, G. Platero$^1$}
\affiliation{1--Instituto de Ciencia de Materiales de Madrid, CSIC, Cantoblanco, 28049 Madrid, Spain\\
2--National Research Council Canada, 1200 Montreal Road., Ottawa, Ontario K1A 0R6, Canada\\
3--D\'epartement de physique, Universit\'e de Sherbrooke, Sherbrooke, Quebec J1K 2R1, Canada}

\begin{abstract}
Tunneling in a quantum coherent structure is not restricted to only nearest neighbours. Hopping between distant sites is possible via the virtual occupation of otherwise avoided intermediate states. Here we report the observation of long-range transitions in the transport through three quantum dots coupled in series. A single electron is delocalized between the left and right quantum dots, while the centre one remains always empty. Superpositions are formed and both charge and spin are exchanged between the outermost dots. The delocalized electron acts as a quantum bus transferring the spin state from one end to the other. Spin selection is enabled by spin correlations. The  process is detected via the observation of narrow resonances which are insensitive to Pauli spin blockade.
\end{abstract}
\pacs{
73.23.Hk, %Electronic transport in mesoscopic systems -- Coulomb blockade; single-electron tunneling.
85.35.Be, %Electronic and magnetic devices; microelectronics -- Quantum well devices (quantum dots, quantum wires, etc.)
73.63.Kv%Electronic transport in nanoscale materials and structures -- Quantum dots
}
\maketitle

Superpositions of indirectly coupled states are possible in quantum mechanics even when the intermediate states are far apart in energy. This is achieved via higher-order transitions in which the energetically forbidden intermediate states are only virtually occupied. Such long-range transitions lie at the core of the theory of the chemical bond~\cite{pauling} and are present in chemical reactions~\cite{gray}, solid state spin phenomena~\cite{anderson}, quantum optics~\cite{arimondo}, and even biological processes~\cite{lambert}. Interest in such phenomena has increased recently within the context of quantum information processing with the possibility of a low dissipation transfer of quantum states~\cite{greentree} or a coherent manipulation of two distant qubits~\cite{kloeffel}.

Semiconductor quantum dot arrays provide a fully tunable platform for manipulating the coherent coupling of quantum states. Great control has already been demonstrated in the double quantum dot system with the observation of molecular-like superpositions via clear resonances in the current flowing through the system~\cite{vanderwiel}. The spin degree of freedom plays a critical role and has led to various proposals utilizing quantum dots as spin or coded spin qubits~\cite{hanson}. An extension to fully coherent triple quantum dot circuits has recently been achieved~\cite{gaudreau,schroer,rogge,ghislain,amaha-spin,gaudreau-qubit,laird}. In addition to being a first step towards more complex quantum simulation architectures~\cite{barthelemy}, such devices make it possible to investigate phenomena which rely on quantum superpositions of distant states mediated by tunneling~\cite{saraga,superexchange}. Long-range tunneling involves the transfer of states from one side of the three-dot array to the other without the occupation of the centre site. A recent experiment reported the observation of such an effect as a transport resonance~\cite{maria}: if the two edge dots of the triple quantum dot array are coupled to source and drain electron reservoirs, left-right superpositions provide a direct channel for the current. The relevant resonant transitions can be measured by time-resolved charge detection~\cite{braakman}. Similar phenomena can also be invoked for the formation of resonant-hybrid states in triangular quantum dot configurations~\cite{amaha}.% or  time resolved charge detection~\cite{braakman}.

In the previous experiments~\cite{maria,braakman} long-range charge transfer required transitions of two electrons:
The electron in the middle dot was exchanged and its spin was potentially flipped during the process. In this Letter we report a remarkably different situation involving the experimental observation of long-range single-electron tunneling resonances for which a simple model confirms that the centre dot is only ever virtually occupied. Hence, the spin of the tunneled electron is well defined.

\begin{figure}[t]
\begin{center}
\includegraphics[width=\linewidth,clip] {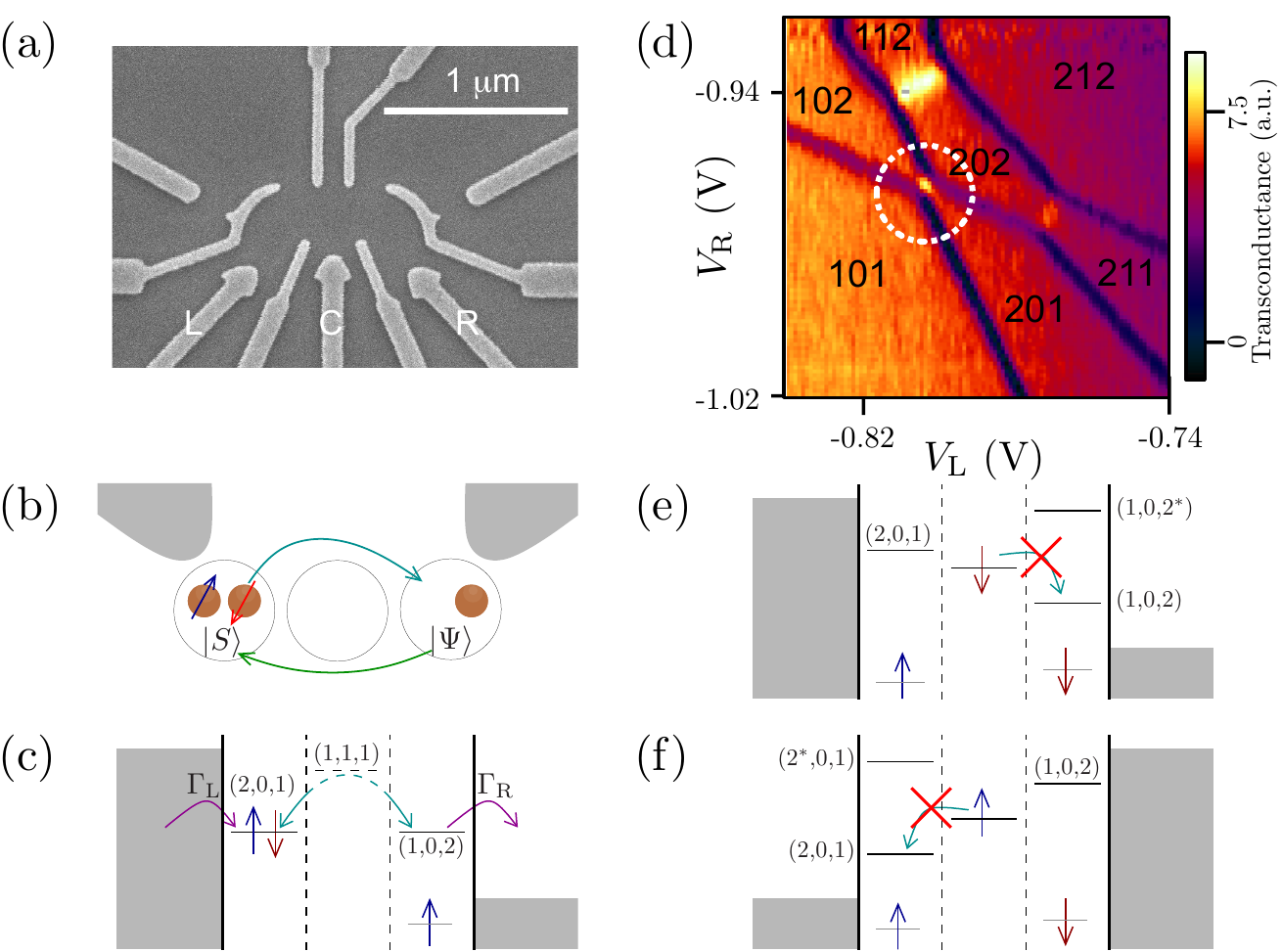}
\end{center}
\caption{\label{scheme} (a) Scanning electron micrograph of a device similar to the one used in the experiments.  (b),© Schematic description of the L-R resonance. A singlet in one of the dots allows for the long-range transfer of an arbitrary spin in the opposite dot.  (d) Zero bias stability diagram of the triple quantum dot transconductance from Ref.~\cite{ghislain} as measured with the left charge detector at a fixed C gate voltage while varying the left (horizontal) and right (vertical) gate voltages. The size of the (1,1,1) region is tuned with the C gate voltage so it closes to a point, while the (2,0,2) region grows. The electronic configurations involved in QP1 are (1,0,1), (2,0,1), (1,1,1), and (1,0,2) and in QP2 (2,0,1), (1,1,1), (1,0,2), and (2,0,2)~\cite{ghislain}.   (e),(f) Schematic description of spin blockade in the forward and backward bias direction, respectively \cite{maria}. 
}
\end{figure}

We investigate the resonance between ($N_\text{L}$,$N_\text{C}$,$N_\text{R}$)=(2,0,1) and (1,0,2) configurations, where $N_l$ is the number of electrons in each dot. Because of the conservation of the total spin, the long-range charge tunneling is necessarily accompanied by a long-range transfer of spin: Of the two electrons forming a singlet in one of the dots, one is transferred to the other edge dot, while the one left behind necessarily has the same spin $\sigma$ as the initially unpaired electron: $|{\uparrow}{\downarrow},0,\sigma\rangle\leftrightarrow|\sigma,0,{\uparrow}{\downarrow}\rangle$, cf. Fig.~\ref{scheme}(b) and (c).

We therefore explore this process where two to four electrons occupy the device and importantly where the centre dot remains empty throughout. Two quadruple points (QP1 and 2) appear in the stability diagram at the position where configurations (2,0,1) and (1,0,2) are close to degeneracy with either (1,0,1) or (2,0,2)~\cite{ghislain}, cf. Fig.~\ref{scheme}(d)~\cite{cooldown}. The configuration (1,1,1) serves as an intermediate state for transport. These configurations have been used to demonstrate Landau-Zener-St\"uckelberg oscillations~\cite{gaudreau-qubit}, the exchange-based qubit~\cite{studenikin} as well as the resonant-exchange qubit~\cite{medford}. In transport, this region is affected by the Pauli exclusion principle: Transitions from (1,1,1) into either (2,0,1) or (1,0,2) are forbidden whenever the electrons in the centre and in the singularly occupied edge dots have the same spin. This effect, known as spin blockade in double quantum dots~\cite{ono}, becomes bipolar in a triple dot~\cite{maria}: Current is blocked by the occupation of triplet states regardless of the applied bias direction [Figs.~\ref{scheme}(e) and~\ref{scheme}(f)]. 

\begin{figure}[t]
\begin{center}
\includegraphics[width=\linewidth,clip] {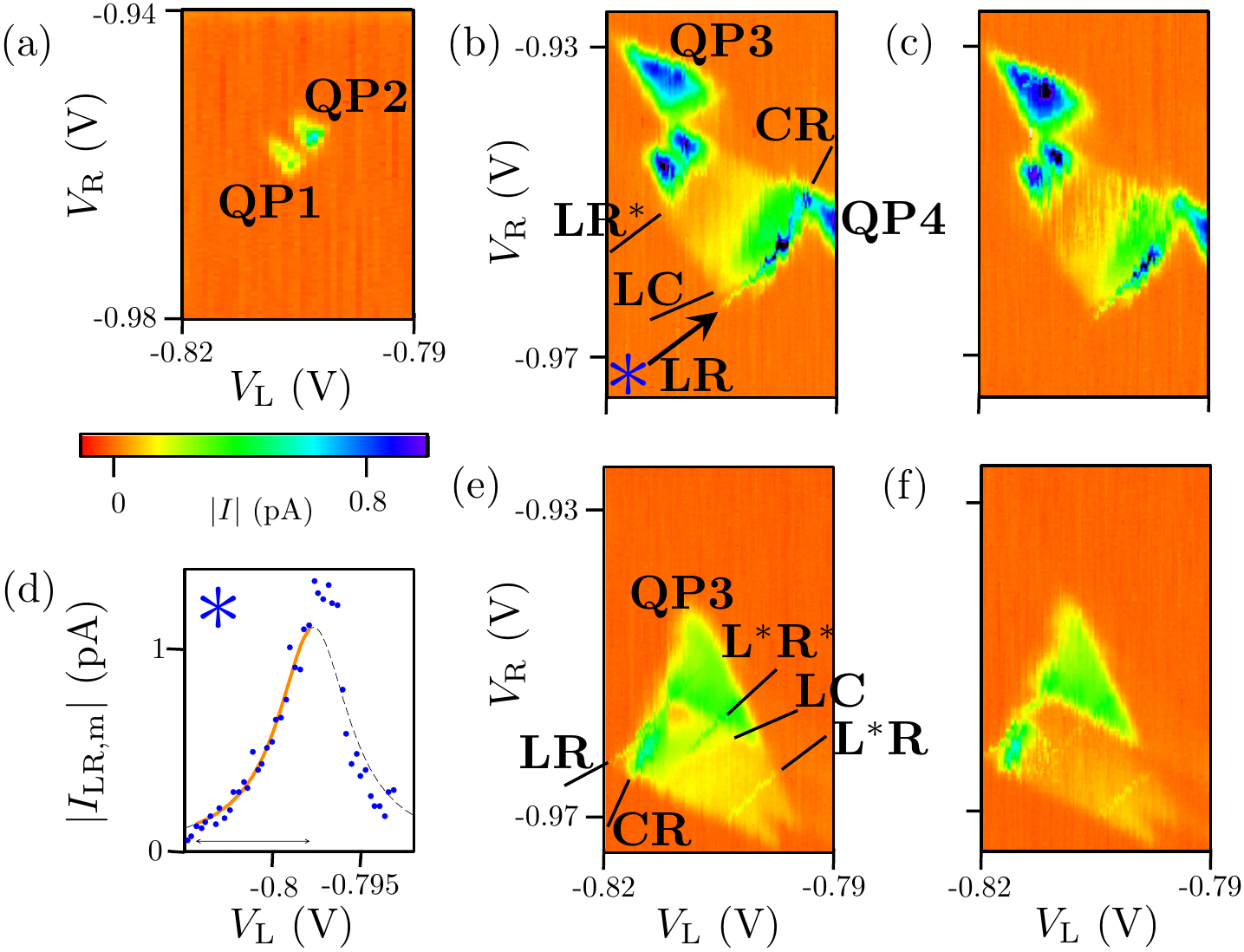}
\end{center}
\caption{\label{exp}  (a) Current through the TQD flows only at two spots (quadruple points 1 and 2 of Ref.~\cite{ghislain}) when the bias is 0.1~mV at magnetic field B=0~T. The electron temperature is $\sim$110~mK. Current at a larger bias of 0.5~mV of
either polarity and in zero or finite parallel magnetic field are in (b)--(f). (b) and $($c) are at 0.5~mV (forward bias), while (e) and (f) are at -0.5~mV (reverse bias). (b) and (e) are at B=0~T, while c) and f) are at B=0.2~T applied parallel to the 2DEG. At 0.5~mV of either polarity, QP1 and 2 expand into two large triangular regions with variable current intensity. QP3 and 4 are also seen as small triangles in forward bias, but QP4 falls outside the maps presented in reverse bias. Resonances are seen in the QP1 and 2 triangles and are labeled according to the dots that are in resonance based on their slopes. Labels with an asterisk mean the dot in question is in an excited state. (d) Current maximum along the forward bias L-R resonance. Its dependence on the detuning with the intermediate state fits well with a model of single-electron virtual tunneling. The orange line marks the region used for the fitting.
}
\end{figure}

{\it Transport measurements.---}The triple quantum dot potential is defined in the two-dimensional electron gas (2DEG) of a GaAs/AlGaAs heterostructure via electrostatic gates. A scanning electron micrograph of the sample is shown in Fig.~\ref{scheme}(a). Important parameters such as the tunnel couplings as well as the energy level spectrum can be tuned by applying appropriate voltages to relevant gates. A \unit[0.5]{meV} bias voltage is applied across the device. This expands the region in the stability diagram where current can flow from the small localized quadruple points [cf. Fig.~\ref{exp}(a)] into larger triangular regions. In Fig.~\ref{exp}, we show the current spectroscopy as a function of the gate voltages, $V_\text{L}$ and $V_\text{R}$, applied to the left and right plunger gates, respectively. Resonance lines can be observed in the triangular regions. Their origin can be identified by comparing their slope to charge transfer lines in the low bias stability diagram. They are labeled in the figure by the dots in which the charge fluctuates: L-C  and C-R lines correspond to the resonance of (1,1,1) with (2,0,1) and (1,0,2) states, respectively. These processes are affected by spin blockade, as evidenced by their dependence on a weak magnetic field: At zero magnetic field [Figs.~\ref{exp}(b) and~\ref{exp}(e)], current leaks due to spin relaxation processes mediated by the hyperfine interaction with nuclei of the host material~\cite{koppens}. At \unit[0.2]{T} [Figs.~\ref{exp}(c) and~\ref{exp}(f)], the finite Zeeman splitting considerably reduces the spin relaxation rate and hence spin blockade persists in both bias polarities.
%Notably different from what is observed in double quantum dots --where spin blockade appears only in one bias direction\cite{ono}--, here spin blockade is present in both bias polarities. The triple dot becomes a spinsulator~\cite{maria}.

Most significantly for this work, L-R lines are observed when (2,0,1) and (1,0,2) states have the same energy. These states are only indirectly coupled through transitions with the intermediate (1,1,1) state. In spite of the fact that the latter (1,1,1) state is detuned away from the other states and is therefore energetically forbidden, we do observe a sharp resonance [Fig.~\ref{exp}(b)--\ref{exp}(f)]. We interpret these lines in terms of the delocalization of one electron between the left and right dots mediated via the virtual occupation of the centre one~\cite{cotunnel}. 
These higher-order transitions depend on the effective hopping $\tau_\text{LC}\tau_\text{CR}/\Delta E$, where $\tau_{ij}$ are the interdot couplings~\cite{ratner}. 
They are modulated by the detuning of states (2,0,1) and (1,0,2) from (1,1,1), $\Delta E(V_\text{L},V_\text{R})$. 
The expected resonance height depends on detuning as $I_\text{LR,m}\propto[1+\alpha\Delta E^2]^{-1}$, in excellent agreement with the experiment, as shown in Fig.~\ref{exp}(d).
%, through the effective hopping $\tau_\text{LC}\tau_\text{CR}/\Delta E$, where $\tau_{ij}$ are the interdot couplings~\cite{ratner}. 
%$\Delta E(V_\text{L},V_\text{R})$ is the detuning of states (2,0,1) and (1,0,2) with (1,1,1).
%The expected dependence of the maximum current along the L-R line on the detuning of states (2,0,1) and (1,0,2) with (1,1,1), $\Delta E(V_\text{L},V_\text{R})$, 
%The dependence of the maximum current along the L-R line on the detuning is well reproduced by a single-electron model of virtual tunneling: $I_\text{LR,m}\propto[1+\alpha\Delta E^2]^{-1}$, as shown in Fig.~\ref{exp}(d).
The spin of the electron that tunnels from the doubly occupied dot is selected by the spin in the other singly occupied edge dot. Thus, L-R lines are not spin blockaded and survive the application of a magnetic field, see Fig.~\ref{exp}. Note the presence of several sets of L-R resonances: one of which involves only singlet states while the others are due to tunneling into an excited state, so that triplet states may be formed in one or both dots (which we label as L-R$^*$, L$^*$-R or L$^*$-R$^*$ lines). 

{\it Theoretical model.---}In order to better understand the relevant processes, we consider a simple model consisting of three Anderson impurities tunnel coupled in series and to two fermionic reservoirs: $\hat{H} = \hat{H}_{\text{TQD}}+\hat{H}_{\text{B}}+\hat{H}_{\text{coupl}}+\hat{H}_{\text{lead}}$, with the triple quantum dot described by
\begin{eqnarray}
\label{ham}
\hat{H}_{\text{TQD}} &=& \sum_{ik\sigma}\varepsilon_{ik}\hat{c}^{\dagger}_{ik\sigma}\hat{c}_{ik\sigma}+\sum_{i}U_{i}\hat{n}_{i\uparrow}\hat{n}_{i\downarrow}+\sum_{i\neq j}\frac{U_{ij}}{2}\hat{n}_{i}\hat{n}_{j}\nonumber\\
&&
{+}\sum_{i}J{\bf S}_{i0}{\bf S}_{i1}{-}\sum_{i\neq j,k,\sigma}{\tau}_{ij}(\hat{c}^{\dagger}_{ik\sigma}\hat{c}_{jk\sigma}{+}\text{H.c.}),
\end{eqnarray}
$\hat{H}_{\text{lead}} = \sum_{lq\sigma}\varepsilon_{lq\sigma}\hat{d}^{\dagger}_{lq\sigma}
\hat{d}_{lq\sigma}$ describes the leads $l=\text{L,R}$, and $\hat{H}_{\text{coupl}}=\sum_{lqk\sigma}\gamma_{l}(\hat{d}^{\dagger}_{lq\sigma}\hat{c}_{lk\sigma}+\text{H.c.})$ represents the dot-lead tunneling coupling. $\varepsilon_{ik}$ is the on site energy in dot $i$, where $k$=0,1 accounts for the ground and excited states, $U_i$, $U_{ij}$ are the onsite and interdot Coulomb interactions, respectively, $J$ is an exchange term and $\tau_{ij}$ is the interdot hopping. We also include a term to account for the magnetic field: $\hat{H}_{\text{B}} =\sum_{ik}\Delta_{i}\hat{S}_{ik}^z$, where the Zeeman splittings $\Delta_i$ are inhomogeneous along the system. This is the case in the presence of the inhomogeneous hyperfine interaction, which results in a different Overhauser field in each dot. We assume a weak device-lead coupling and derive the quantum master equation $\dot\rho={\cal L}\rho$ for the reduced density matrix of the triple quantum dot, $\rho$~\cite{maria}. From its stationary solution,  we obtain the current $I=e{\cal J}\rho$, where the current operator is given by the tunneling rates $\Gamma_l=2\pi\nu_l|\gamma_l|^2/\hbar$, where $\nu_l$ is the density of states in the leads. We assume that spin flip transitions are two orders of magnitude slower than $\Gamma$ and introduce them by using phenomenological rates. To account for their magnetic field dependence, we assume a Lorentzian dependence on the Zeeman splitting: Spins relax faster at low magnetic fields~\cite{koppens}.

\begin{figure}[t]
\begin{center}
\includegraphics[width=0.9\linewidth,clip] {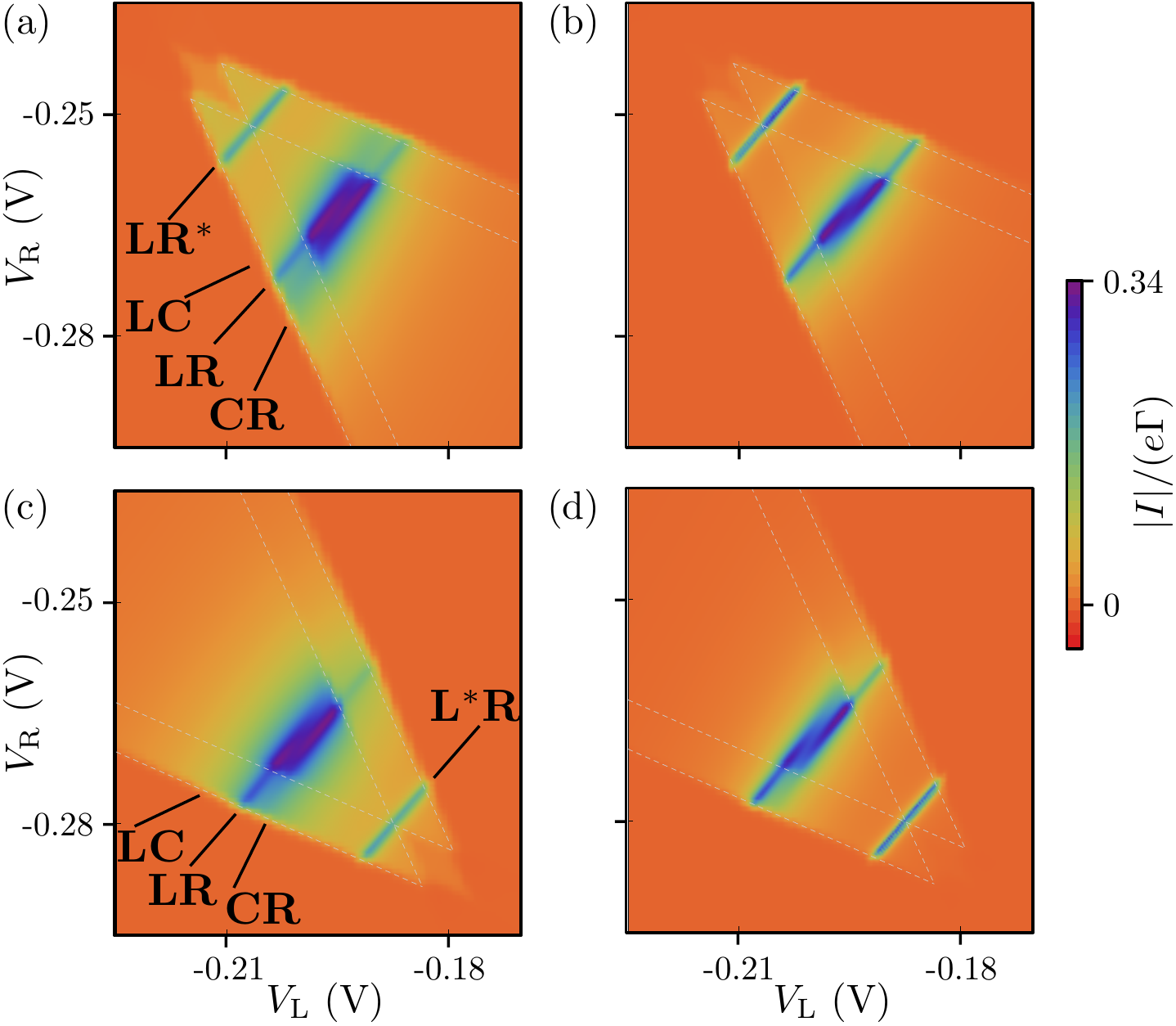}
\end{center}
\caption{\label{theory} Calculated current through the triple quantum dot for a bias $|\mu_L-\mu_R|=\unit[0.5]{meV}$ applied in the (a)-(b) forward and (c)-(d) backward direction. (a) and (c): Leakage currents due to spin flip processes at zero magnetic field reveal the position of the different resonance lines. (b) and (d): Under the application of a magnetic field of 0.2 T, spin blockade is enhanced which affects the L-C and C-R resonances but not the L-R lines.   We only consider states participating in QP1 and 2. White dashed lines mark the crossing of the dots and lead chemical potentials. Parameters: $T=\unit[110]{mK}$ and (in meV): $\varepsilon_{i1}-\varepsilon_{i0}=0.33$, $J=-0.011$, $\tau_{LC}=0.022$, $\tau_{CR}=0.015$, $U_L=1.72$, $U_R=1.22$, $U_{LC}=U_{CR}=0.28$, $U_{LR}=0.167$, $\Gamma_L=2.8\times10^{-3}$, $\Gamma_R=2.1\times10^{-3}$.
}
\end{figure}

We focus on the dynamics of the quadruple points where (2,0,1)$\leftrightarrow$(1,0,2) transitions are important. We therefore neglect the influence of the additional quadruple points visible in the experiment where L-R resonances are not possible. As can be seen in Fig.~\ref{theory}, our minimal model reproduces the main features of the experiment~\cite{repr}. At zero magnetic field, a leakage current flows in the region delimited by the conditions $E(2,0,n)-E(1,0,n)=\mu_\text{L}$ and $E(n,0,2)-E(n,0,1)=\mu_\text{R}$ for the chemical potentials of the two leads, where $n$=1,2 for each triangle (marked as pale dashed lines in Fig.~\ref{theory}). The current is dominated by the degeneracy point of (2,0,1), (1,1,1) and (1,0,2) charge configurations in the region where the two triangles overlap. As a result, the different resonance lines cross. The coherent character of interdot tunneling results in an anticrossing which depends on the asymmetry $\tau_\text{LC}/\tau_\text{CR}$; cf. Figs.~\ref{theory}(a) and~\ref{theory}(c).
A second L-R line appears for each bias polarity due to a tunneling to excited states in the dot coupled to the collector: L-R$^*$ (forward bias) and L$^*$-R (backward bias). At the finite magnetic field [cf. Figs.~\ref{theory}(b) and~\ref{theory}(d)], spin blockade is much more efficient and current is strongly suppressed, with only the L-R transitions contributing to the current.

Note that L-R resonances are much narrower than all the other resonances. This observation is due to the different nature of the electron transfer process: For nearest neighbour transitions, such as those taking place along L-C lines, tunneling is dominated by a sequence of two hoppings, $\tau_{LC}$ and $\tau_{CR}$. It is very different for the L-R resonances, where tunneling depends on the effective coupling $\tau_\text{LC}\tau_\text{CR}/\Delta E$.
%to the centre dot is energetically forbidden. 
%These higher order transitions depend to lowest order~\cite{ratner} on the coupling $\tau_\text{LC}\tau_\text{CR}/\Delta E$.

We emphasize that the centre dot is kept empty during the transfer of charge from the left to the right lead: (1,0,1)$\rightarrow$(2,0,1)$\leftrightarrow$(1,0,2)$\rightarrow$(1,0,1) at forward bias (and vice versa at backward bias). This is not only counterintuitive but has some additional advantages. On one hand, thanks to the double occupancy of the source dot, spin blockade is avoided. On the other hand, out of the three electrons in the initial state, there is only one that can be transferred across the system. Thus, it is clear that L-R lines consist of the transfer of a single electron with a well-defined spin directly from the left to the right dot. The observation of the sharp resonance lines confirms that the interdot tunneling is coherent and that the sign of the spin is conserved in the transfer process. This is an important observation for the future application of such processes for spin bus applications. This was not the case of previous experiments~\cite{maria,braakman} where the transfer of charge from one side to the other could potentially be accompanied by a spin flip (i.e., spin states were mixed).

\begin{figure}[t]
\begin{center}
\includegraphics[width=\linewidth,clip] {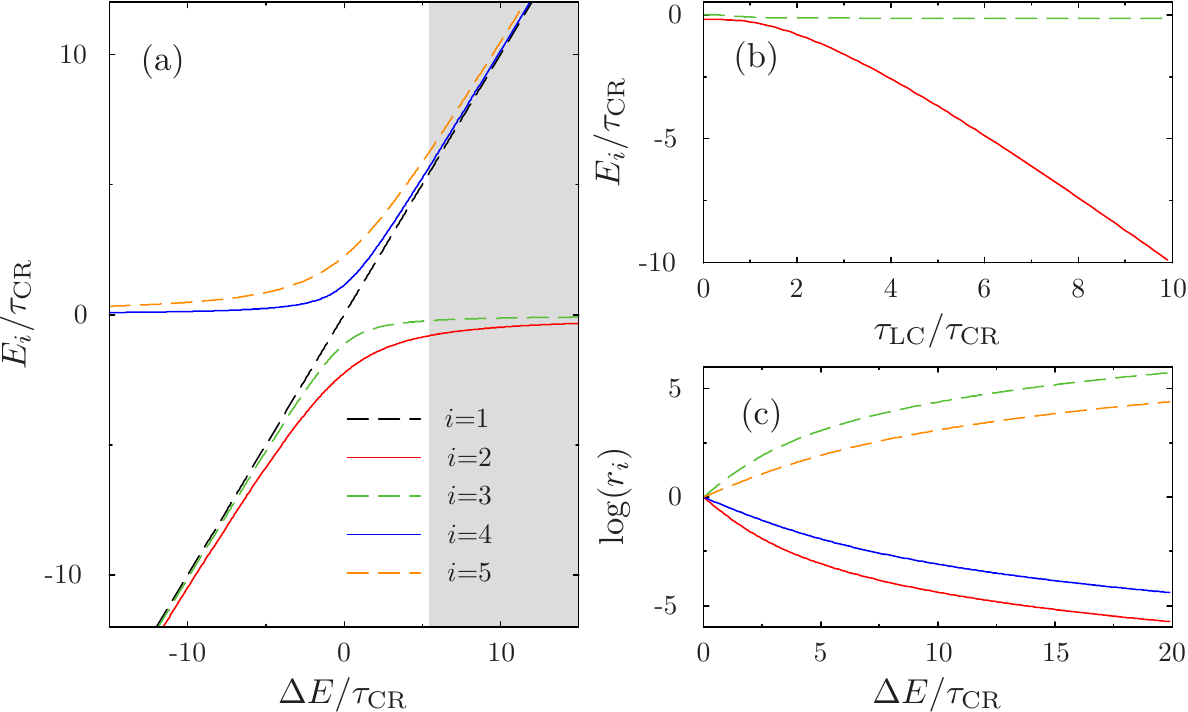}
\end{center}
\caption{\label{eigen} Spectrum at the L-R resonance for the configuration in Fig.~\ref{theory}, with fixed $\tau_\text{CR}$. (a) The eigenergies $E_i$ as a function of the detuning with the intermediate states, $\Delta E$. The grey shadow marks the region of interest where only ${\bf v}_2$ contributes to transport (see text). (b) The eigenenergies $E_2$ and $E_3$ split with the interdot hopping inhomogenity, here for $\Delta E=10\tau_\text{CR}$. (c) Contribution of the (1,1,1) states to the eigenstates ${\bf v}_i$ parametrized by the factor $r_i=\langle\text{v}_i|{\cal P}|\text{v}_i\rangle/\langle\text{v}_i|(1-{\cal P})|\text{v}_i\rangle$, with the projector ${\cal P}=\sum_j|C_j\rangle\langle C_j|$. We do not plot $r_1$ as it exactly diverges.
}
\end{figure}

We gain further insight into the process by looking at the eigenstates of the triple quantum dot system. We identify their role in the process and how to control and optimize it. Let us consider the subspace with three electrons: two with spin $\sigma$ and one with opposite spin $\bar\sigma$. The L-R transition can only connect the states $|\text{L}\rangle=|{\uparrow}{\downarrow},0,{\sigma}\rangle$ and $|\text{R}\rangle=|{\sigma},0,{\uparrow}{\downarrow}\rangle$, with the intermediate states being $|\text{C}_1\rangle=|{\sigma},{\bar\sigma},{\sigma}\rangle$, $|\text{C}_2\rangle=|{\bar\sigma},{\sigma},{\sigma}\rangle$ and $|\text{C}_3\rangle=|{\sigma},{\sigma},{\bar\sigma}\rangle$. Let us take, for simplicity, the case of symmetric interdot couplings: $\tau_{LC}=\tau_{CR}=\tau$. We can diagonalize the corresponding block of $\hat{H}_{\text{TQD}}$, whose unnormalized eigenstates read:
%${\bf v}_1=\sum_i|\text{C}_i\rangle/\sqrt{3}$, ${\bf v}_{2,3}=[|\text{L}\rangle-|\text{R}\rangle+\beta_\pm(2|\text{C}_1\rangle-|\text{C}_2\rangle-|\text{C}_3\rangle)]/\sqrt{2+4\beta_\pm^2}$, and ${\bf v}_{4,5}=[|\text{L}\rangle+|\text{R}\rangle-\alpha\pm(|\text{C}_2\rangle-|\text{C}_3\rangle)]/\sqrt{2+2\alpha_\pm^2}$, with $\alpha_{\pm}=2\tau/(\Delta E\pm\sqrt{\Delta E^2+4\tau^2})$ and $\beta_{\pm}=2\tau/(\Delta E\pm\sqrt{\Delta E^2+12\tau^2})$.
${\bf v}_1=\sum_i|\text{C}_i\rangle$, ${\bf v}_{2,3}=|\text{L}\rangle-|\text{R}\rangle+\beta_\pm(2|\text{C}_1\rangle-|\text{C}_2\rangle-|\text{C}_3\rangle)$, and ${\bf v}_{4,5}=|\text{L}\rangle+|\text{R}\rangle-\alpha_\pm(|\text{C}_2\rangle-|\text{C}_3\rangle)$, with $\alpha_{\pm}=2\tau/(\Delta E\pm\sqrt{\Delta E^2+4\tau^2})$ and $\beta_{\pm}=2\tau/(\Delta E\pm\sqrt{\Delta E^2+12\tau^2})$. We plot the corresponding eigenvalues in Fig.~\ref{eigen}. For the case of interest, $\Delta E\gg\tau$, we obtain to leading order in a series expansion: ${\bf v}_2\rightarrow|\text{LR}_-\rangle$ and ${\bf v}_4\rightarrow|\text{LR}_+\rangle$, ~\cite{dark}
\be
\label{lrst}
|\text{LR}_\pm\rangle=\frac{1}{\sqrt{2}}(|{\uparrow}{\downarrow},0,{\sigma}\rangle\pm|{\sigma},0,{\uparrow}{\downarrow}\rangle).
\ee
The occupation of these two eigenstates involves the delocalization of an electron between the left and right dots without the participation of the centre one. They are thus responsible for the L-R lines. Furthermore, as an electron tunnels from one extreme of the structure to the other, an arbitrary spin $\sigma$ is transferred in the opposite direction, e.g., $\sum_\sigma c_\sigma{|}{\uparrow}{\downarrow},0,{\sigma}\rangle\rightarrow\sum_\sigma c_\sigma{|}{\sigma},0,{\uparrow}{\downarrow}\rangle$; i.e., the tunneling electron acts as a spin bus. Hence, via this mechanism any spin state prepared in one of the dots, ${|}\psi\rangle_l=c_\uparrow{|}{\uparrow}\rangle_l+c_\downarrow{|}{\downarrow}\rangle_l$, can be transferred to the other edge: ${|}{\uparrow}{\downarrow}\rangle_\text{L}{\otimes}{|}0\rangle_\text{C}{\otimes}{|}{\psi}\rangle_\text{R}\rightarrow{|}{\psi}\rangle_\text{L}{\otimes}{|}0\rangle_\text{C}{\otimes}{|}{\uparrow}{\downarrow}\rangle_\text{R}$. 

Interestingly, at the L-R resonance, ${\bf v}_2$ is the ground state of the system; see Fig.~\ref{eigen}. Thus it will constitute the main channel for transport. Far from the multiple resonance, i.e., for $\Delta E\gg\tau$, it becomes almost degenerate with ${\bf v}_3\sim2|\text{C}_1\rangle-|\text{C}_2\rangle-|\text{C}_3\rangle$, which to leading order is not coupled to the leads and therefore does not contribute to transport. Being very weakly connected to the leads, states ${\bf v}_1$, ${\bf v}_3$, and ${\bf v}_5$ are slow channels~\cite{v5}. Therefore, their occupation will potentially switch off the desired current through ${\bf v}_2$. 
In order to avoid this effect and enhance the observation of the left to right transitions, the splitting between ${\bf v}_2$ and ${\bf v}_3$ can be increased by introducing inhomogeneous tunnel coupling, $\tau_\text{LC}\ne\tau_\text{CR}$, as shown in Fig.~\ref{eigen}(b). The inhomogeneity slightly modifies the coefficients in the eigenvectors, but does not affect their general properties discussed here. 
%In the opposite limit, $\Delta E\ll\tau$, ${\bf v}_5$ becomes $|{\sigma},{\sigma},{\bar\sigma}\rangle-|{\bar\sigma},{\sigma},{\sigma}\rangle$, i.e. it involves long-range spin swap transitions between electrons in the left and right quantum dots, while at the same time being decoupled from the leads.

{\it Conclusions.---}We report measurements involving the long-range tunneling of a single electron in a triple quantum dot structure. We observe sharp transport resonances which involve coherent single-electron transport from one edge dot to the other while avoiding the occupation of the centre dot. An important difference with previous experiments~\cite{maria,braakman} is the fact that the spin of the electron is well defined during the process. Its spin is selected by an electron occupying the other edge dot via the Pauli exclusion principle. The detected process thus enables a protocol for spin bussing, essential for quantum information architectures.

%--------------
We acknowledge discussions with M. Busl and F. Gallego-Marcos. The high-quality 2DEG material was grown by Z.R. Wasilewski. This work was supported by the Spanish MICINN Juan de la Cierva program (R. S.) and MAT2011-24331, the ITN Grant No. 234970 (EU). G. G. acknowledges funding from the National Research Council Canada -- Centre national de la recherche scientifique collaboration and Canadian Institute for Advanced Research.  A. S. S. acknowledges funding from Natural Sciences and Engineering Research Council of Canada and Canadian Institute for Advanced Research.


\begin{thebibliography}{30}
\bibitem{cooldown}
The results from this paper were taken during the same configuration and cooldown as the data in Fig.~\ref{scheme}(d).
\bibitem{cotunnel}
It is interesting to note the resemblance with cotunneling events~\cite{averin}, but with the important difference that in our case the initial and final states are localized states, not immersed in the lead Fermi seas.
\bibitem{repr}
We do not reproduce L$^*$-R$^*$ lines as the inclusion of excited states in both dots has a heavy computational cost. Additional features of the experiment such as the enhanced current at the top of the triangles in forward bias [cf. Figs.~\ref{exp}(b) and (c)] are attributed to the close proximity of QP3 and 4, not included in the model.
\bibitem{dark}
Note the similarity of the superposition $|\text{LR}_-\rangle$ with dark states in optical resonances~\cite{arimondo} which have been predicted to be measurable in transport through double~\cite{brandes,darkbell} and triple~\cite{cpt,mariadark} quantum dot systems, but have so far not been observed.
\end{thebibliography}
\end{document}